# The first peptides: the evolutionary transition between prebiotic amino acids and early proteins

Peter van der Gulik[1], Serge Massar[2], Dimitri Gilis[3], Harry Buhrman[4], Marianne Rooman[3]

[1] Centrum Wiskunde & Informatica, Kruislaan 413, 1098 SJ Amsterdam, the Netherlands

[2] Laboratoire d'Information Quantique, Université Libre de Bruxelles, CP225, avenue Roosevelt 50, 1050 Brussels, Belgium

[3] Unité de Bioinformatique Génomique et Structurale, Université Libre de Bruxelles, CP 165/61, avenue Roosevelt 50, 1050 Brussels, Belgium

[4] Centrum Wiskunde & Informatica and Universiteit of Amsterdam, Kruislaan 413, 1098 SJ Amsterdam, the Netherlands

## Abstract

The issues we attempt to tackle here are what the first peptides did look like when they emerged on the primitive earth, and what simple catalytic activities they fulfilled. We conjecture that the early functional peptides were short (3 to 8 amino acids long), were made of those amino acids, Gly, Ala, Val and Asp, that are abundantly produced in many prebiotic synthesis experiments and observed in meteorites, and that the neutralization of Asp's negative charge is achieved by metal ions. We further assume that some traces of these prebiotic peptides still exist, in the form of active sites in present-day proteins. Searching these proteins for prebiotic peptide candidates led us to identify three main classes of motifs, bound mainly to $Mg^{2+}$ ions: D(F/Y)DGD corresponding to the active site in RNA polymerases, DGD(G/A)D present in some kinds of mutases, and DAKVGDGD in dihydroxyacetone kinase. All three motifs contain a DGD submotif, which is suggested to be the common ancestor of all active peptides. Moreover, all three manipulate phosphate groups, which was probably a very important biological function in the very first stages of life. The statistical significance of our results is supported by the frequency of these motifs in today's proteins, which is three times higher than expected by chance, with a *P-value* of $3 \cdot 10^{-2}$. The implications of our findings in the context of the appearance of life and the possibility of an experimental validation are discussed.

# Introduction

The development of molecular biology over the past half century has transformed the question of how life appeared from the level of philosophical speculations to the level of rigorous scientific enquiry. A number of key ideas have emerged which govern our thinking about this question, among which the synthesis of some amino acids and sugars by prebiotic synthesis (Miller 1987; Ricardo *et al.* 2004), the RNA-world in which RNA catalyses its own duplication (Gilbert 1986; Johnston *et al.* 2001), and the role of lipid vesicles to limit the spatial extent of the cell precursor (Deamer 1985; Chen 2006; Mansy *et al.* 2008). At some point the controlled synthesis of proteins emerged, and proteins then took over many functions, presumably because of their great specificity and efficiency.

In this letter we address an important question, namely what did the first functional peptides and proteins look like when they emerged during very early life? Answering these questions is a difficult and unsolved issue. Indeed, the smallest natural proteins that can take a stable structure by themselves, without interacting with other biomolecules, are composed of around 20 amino acids (aa's). Some synthetic constructs are even smaller: for example, chignolin, a synthetic protein of 10 aa's, has been shown to have a stable structure in water (Honda *et al.* 2004). But it is difficult to imagine that functional proteins 10-20 aa's long suddenly appeared out of the blue. Indeed the sequences of these proteins are very specific, and reliably making functional proteins this long requires an efficient code for the aa sequence, and an efficient translation mechanism from the code into the protein (see the Discussion section).

Here we propose a solution to this "chicken and egg" paradox by suggesting that specific short peptides 3 to 8 aa's long could have served as catalysts during very early life. Longer proteins would then have gradually evolved from these early precursors. The idea that very short peptides could have had a useful role in very early life has already been put forward by Shimizu (Shimizu 1996, 2004, 2007) who showed that single amino acids and dipeptides could slightly enhance the rates of certain chemical reactions. But obviously some intermediate steps are required between Shimizu's dipeptides and the smallest of today's functional enzymes.

An additional constraint on any theory of how the first functional peptides emerged is that these should be composed exclusively or almost exclusively by the aa's that are efficiently produced by prebiotic synthesis. In most prebiotic synthesis experiments (see next section) the most efficiently produced amino acids are Gly, Ala, Val and Asp (or G, A, V and D in one letter code) (Miller 1987). Of these, the first three are neutral, and Asp is negatively charged. At first sight this is problematic: the absence of positive charges compensating the negative Asp's are likely to limit their ability to form stable structures and to carry out a catalytic activity.

We propose to resolve these conundrums by suggesting that the first peptides were composed of short chains of prebiotic aa's bound to (one or more) positively charged metal ions. Supposing the first peptides to be bound to metal ions solves two problems at once. Namely, the metal ion(s) provide(s) an anchor around which the peptide can organize itself, thereby (at least partially) stabilizing its structure, and secondly it provides a positive charge which would be very useful for catalytic activity.

Later, as the coding and translation mechanisms improved, these very first peptides gradually lengthened, thereby improving the efficiency and specificity of their biological activity. We further conjecture that some of these first peptides, composed of prebiotic aa's bound to metal ions, have been conserved across evolution. The fact that active sites are better conserved than all other protein regions speaks in favor of this conjecture. If this idea is correct, it should be possible to find today the memory of the very first functional peptides in the active sites of some present-day proteins.

The idea of finding in present-day proteins traces of very early life is not entirely new. For instance it has been argued that today's aa abundances reflect the order in which they were introduced in the genetic code (see Zuckerkandl *et al.*, 1971; Jordan *et al.,* 2005, and the criticism of Hurst *et al.*, 2006).

To explore the validity of our conjectures, we carried out a search in the protein structure database to identify all active sites composed almost exclusively of the abundant prebiotic aa's bound to metal ions. In the following we review the abundances of amino acids produced in prebiotic synthesis experiments, then present our precise search methodology, and discuss the classes of prebiotic peptide candidates revealed by the search. These candidates all correspond to active sites in their host protein, and perform catalytic functions likely to be important during very early life. Furthermore there is strong evidence that some of these active sites have been conserved at least since the Last Universal Common Ancestor (LUCA) 3.5 billion years ago. We conclude this letter by discussing in more detail how our suggestions fit into the wider picture of the appearance of life, and how they could be tested experimentally.

## State of the art: Prebiotic amino acids

In this section we review the relative abundances of aa's that are produced in prebiotic synthesis experiments such as those originally designed by Miller (1955), or that are observed in meteorites, in order to determine the most plausible composition of the first peptides that appeared on the primitive earth. These aa abundances are summarized in Table 1. Note that we do not review the determinations of aa's in the interstellar medium, as this issue is under discussion (Kuan *et al.* 2003; Snyder *et al.* 2005; Jones et al. 2007; Zaia *et al.* 2008).

Miller and coworkers performed synthesis experiments in several environments attempting to mimic possible prebiotic circumstances, as the real environment is not agreed upon. The first three environments considered by Miller (1955), and recently reanalyzed by Johnson *et al.* (2008), consist of a highly reducing mixture of $CH_4$, $NH_3$, $H_2O$ and $H_2$, which is sparked in three different ways. They all produced Gly, Ala, Asp, Glu and Ser in significant amounts. Moreover, in the experimental set-up which is nowadays considered the most interesting one (see Johnson et al., 2008), Val was produced in about the same amount as Glu. In later work of Miller and coworkers (Ring *et al.*, 1972), ammonia was only present in trace amounts and the major source of nitrogen was $N_2$. In this case ten proteinaceous aa's were produced, in order of abundance: Ala, Gly, Asp, Val, Leu, Glu, Ser, Ile, Pro and Thr.

Schlesinger & Miller (1983) investigated the effect of changing the carbon source ($CH_4$, CO, or $CO_2$). They found that with CO or $CO_2$ as carbon source, Gly was almost the only amino acid found, with small amounts of Ala and Ser, and trace amounts of Asp and Glu, and, in a single case, Val. Recently, however, Miller and coworkers showed, contrary to previous reports, significant production of Ala, Ser and Glu in neutral atmospheres (Cleaves *et al.*, 2008). They conjectured that nitrite and nitrate, which are also produced by spark discharge in neutral atmospheres, destroyed the amino acids in previous work. Addition of pH buffer and oxidation inhibitor prevents this destruction. They concluded that neutral atmospheres may have been productive in prebiotic synthesis of amino acids, provided the early oceans were buffered sufficiently with respect to pH and redox balance.

Another series of experiments, performed by Plankensteiner *et al.* (2006), assumed the prebiotic atmosphere to be neutral and composed of $CO_2$, $N_2$, $H_2O$. The aa's that were systematically produced were Gly, Ala and Val. In addition, Ser, Pro, and Lys were detected in significant amounts in several experiments. Surprisingly, Asp and Glu were not detected.

An interesting observation is the similarity between the proteinaceous aa's produced in prebiotic synthesis experiments and those present in carbonaceous meteorites, as first noticed by Kvenvolden (1974) and Miller (1974). A detailed analysis of the composition of several meteorites was later performed by Botta *et al.* (2002), and showed the presence of Gly, Ala, Asp, Val, Ser and Glu.

Clearly, some discrepancies occur among the amino acids produced, according to the experimental setups, the environments meant to represent the primitive earth, or the type of meteorites. However, general tendencies are noted: Gly and Ala always appear as the most abundant aa's, and Asp, Val, Ser, and Glu often appear in significant amounts.

The choice we made in this paper of which were the abundant prebiotic aa's was motivated by the prebiotic synthesis experiments performed by Ring *et al.* (1972). These authors found ten proteinaceous aa's as well as several non-proteinaceous ones in detectable amounts, with a spectrum showing strong similarities with that of the Murchison meteorite. In this experiment, Gly, Ala, Asp and Val were the four most abundant aa's, and we chose them as the presumed components of the first active peptides.

## Methods: Search for traces of prebiotic peptides in present-day proteins

We searched for traces of prebiotic peptides in the RCSB Protein structure DataBase (PDB) (Berman *et al.* 2002). For the purpose of selecting all PDB entries that contain a protein structure interacting with one metal ion at least and of identifying all interactions between ions and aa's, we used the PDBsum summary and analysis tool (Laskowski *et al.* 2005). The metals considered are Li, Be, Na, Mg, K, Ca, Cr, Mn, Fe, Co, Ni, Cu, Zn, and Se.

The prebiotic peptide candidates we searched for in the PDB were assumed to be built almost exclusively from the aa's that were presumably the most abundant in early times, *i.e.* Gly, Ala, Val and Asp (see previous section). Moreover, these peptide candidates were required to be bound to metal ions, present in the prebiotic environment, in order to neutralize the negative charge of Asp and to allow the formation of well defined tertiary structures and catalytic functions.

The exact criteria that we used to characterize prebiotic peptide candidates are the following. Firstly, these peptides must correspond to sequence regions of maximum length 8, the longest prebiotic peptides envisaged. Note that doubling this maximum length did not change the results of the search. Secondly, they must be bound to a metal ion, and this ion may not interact with any other part of the protein. Thirdly, the ion must be bound to the abundant prebiotic aa's only, among which three or more Asp residues. Requiring three Asp's at least increases the strength and specificity of the peptide-ion binding and thus its chance to be part of the protein active site in addition to stabilizing the structure. Indeed, when relaxing the criterion and accepting two Asp's bound to an ion, many spurious fragments were selected. Fourthly, the sequence starting at the first and ending at the last residue bound to the ion, and including start and end residues, should be composed of the abundant prebiotic amino acids for 80% at least. Lastly, if this sequence is flanked in the host protein by abundant prebiotic aa's, these aa's are added to the peptide.

To estimate the statistical significance of finding these prebiotic peptide candidates in the PDB, we computed *E*, the number of such peptides expected to occur by chance, as follows. Let $E_{ij}$ be the expected number of sequences of length $i$ with $j$ residues binding an ion. We compute this as $E_{ij} = P_{ij} N_{ij}$, where $P_{ij}$ is the probability that a motif matches our prebiotic criteria, and $N_{ij}$ is the number of motifs of length $i$ with $j$ residues attached to the ion present in the PDB. The probabilities $P_{ij}$ are estimated as:

$$P_{ij} = \binom{i-2}{j-2} \sum_{k=3}^{j} \binom{j}{k} P(D|ion)^k P(A/G/V|ion)^{j-k}$$
$$\sum_{m=\lceil 0.8i \rceil - j}^{i-j} \binom{i-j}{m} P(D/A/G/V|\overline{ion})^m \overline{P(D/A/G/V|ion)}^{i-j-m} \quad, \quad (1)$$

where '/' means 'or', and $P(y|ion)$ [$P(y|\overline{ion})$] is the fraction of aa's of type $y$, given that they are bound [not bound] to one of the considered ions; these fractions are computed from the full PDB. To obtain this formula, we first fixed the outer aa's binding the ion, which leaves $\binom{i-2}{j-2}$ possibilities to choose the location of the remaining $j$-2 prebiotic aa's that also bind the ion. Among these we have to choose at least three Asp's, giving rise to the sum from k=3 to j. Last we have to choose, out of the $i$-$j$ remaining positions that do not bind the ion, at least $\lceil 0.8i \rceil - j$ prebiotic aa's in order to have at least 80% such aa's in the motif. The total expected number of prebiotic motifs is given by:

$$E = \sum_{i=3}^{8} \sum_{j=3}^{i} E_{ij} = \sum_{i=3}^{8} \sum_{j=3}^{i} P_{ij} N_{ij} \quad . \quad (2)$$

However, this formula contains several overcountings. First, the PDB contains many highly similar, homologous, proteins. This implies that when a motif is found in one of them, it is almost automatically found in the others. To correct for this bias, we replace $N_{ij}$ by $N_{ij}^{cor}$, which counts only the motifs that have at least one different aa, or that have at least 20% different aa's in a sequence stretch of 10 residues around the middle residue of the motif. Another correction is due to the fact that two consecutive Asp's binding an ion are very rarely observed. Indeed, the joint probability of finding two Asp's bound to an ion at consecutive positions along the sequence is much lower than the product of the independent probabilities of finding an Asp bound to an ion; the ratio of these probabilities is only equal to 0.05. This effect is due to geometrical constraints imposed by the polypeptide chain, and the unfavorableness of having two negative charges close together. This effect is also, but to a lower extent, observed for other consecutive aa's bound to an ion. To take this effect into account, we have to distinguish between the different ion-binding motifs, which we label by the index $v$. For example, for a motif of length $i=5$ with $j=3$ residues bound to an ion, the $\binom{i-2}{j-2}=3$ possible motifs are `iixxi`, `ixixi` and `ixxii`, where `i` and `x` denote aa's bound and not bound to the ion, respectively; $N_{ijv}^{cor}$ with $i=5, j=3$ and $v=1$ corresponds thus to the number of non-homologous matches of motifs of type `iixxi` in the PDB. This leads us to the corrected expected value $E^{cor}$:

$$E^{cor} = \sum_{i=3}^{8} \sum_{j=3}^{i} E_{ij}^{cor} = \sum_{i=3}^{8} \sum_{j=3}^{i} \sum_{v=1}^{\alpha_{ij}} P_{ijv}^{cor} N_{ijv}^{cor} \quad \text{with} \quad \alpha_{ij} = \binom{i-2}{j-2} \quad, \quad (3)$$

where the corrected probability $P_{ijv}^{cor}$ is given by:

$$P_{ijv}^{cor} = \sum_{k=3}^{j}\sum_{w=1}^{\beta_{jk}} P(D|D,ion)^{n_{vw}^{DD}} P(D|a,ion)^{n_{vw}^{Da}} P(D|ion)^{k-n_{vw}^{DD}-n_{vw}^{Da}} P(a|D,ion)^{n_{vw}^{aD}} P(a|a,ion)^{n_{vw}^{aa}} P(a|ion)^{j-k-n_{vw}^{aD}-n_{vw}^{aa}}$$

$$\sum_{m=\lceil 0.8i \rceil - j}^{i-j} \binom{i-j}{m} P(D/a|\overline{ion})^m P(\overline{D/a}|\overline{ion})^{i-j-m} \quad , \quad \text{with} \quad \beta_{jk} = \binom{j}{k} \quad . \quad (4)$$

where $a=A/G/V$, and $P(y|z,ion)$ is the probability that an aa bound to an ion is of type $y$, given that the next aa along the sequence is bound to the same ion and is of type $z$. The index $w$ labels the $\beta_{jk}$ motifs of type $v$; for example, for $j=4$ and $k=3$, the $\beta_{jk}=4$ ion-binding motifs of type `iiixi` are `DDDxa`, `DDaxD`, `DaDxD` and `aDDxD`. The integers $n_{vw}^{yz}$ denote the number of times, in a motif of type $v$ (*e.g.* `iiixi`) and subtype $w$ (*e.g.* `DDaxD`), two successive aa's are bound to an ion, the first being of type $y$ and the second of type $z$. In the example `DDaxD`, $n_{vw}^{DD}=1$, $n_{vw}^{Da}=1$, $n_{vw}^{aD}=0$ and $n_{vw}^{aa}=0$.

To objectively compare $E$ (or $E^{cor}$), the expected number of prebiotic peptide candidates, and $N$ (or $N^{cor}$), their actual number observed in the PDB, we estimate the *P*-value, defined as the probability of finding at least $N$ (or $N^{cor}$) prebiotic peptide candidates if the null hypothesis was true. For this, we assume that $E$ (or $E^{cor}$) follow a Poisson distribution, as it is a discrete probability distribution only defined for positive values. The *P*-value is then given by:

$$P\text{-value} = \sum_{k=N}^{\infty} \frac{E^k e^{-E}}{k!} \quad . \quad (5)$$

## Results: Prebiotic peptide candidates

Using the above definition of prebiotic peptide candidates, we screened systematically the whole PDB, and found two different ion binding motifs: DxDxD and DxxxxDxD, where x stands for any aa. These two motifs differ in the relative positions along the sequence of the three Asp residues binding the ion. The first binding motif further divides into four groups, according to the aa's between and flanking the Asp's, as summarized in Table 2.

The first two DxDxD binding motifs correspond to active sites of RNA polymerases, and are both bound to $Mg^{2+}$, or sometimes $Mn^{2+}$ ions. The first motif, with strictly conserved sequence ADFDGD, or even NADFDGD if one includes the non abundant prebiotic residue N, is found in DNA-dependent RNA polymerase, the normal RNA polymerase engaged in protein-coding gene expression. More precisely, it is found in the second domain of RNA polymerase Rpb1, using the nomenclature of the protein families database Pfam (Finn *et al.* 2008).

It is worthwhile to linger a few moments on the age of this NADFDGD sequence. It corresponds to the active site in the mRNA-producing enzyme of many organisms, such as *Homo sapiens*, the nematode worm *Caenorhabditis elegans,* but also in the yeast *Saccharomyces cerevisiae* and the plant *Arabidopsis thaliana* (Iyer *et al.*, 2003). This sequence seems thus to be present in all eukaryotes, which populate the earth since two billion years at least. Moreover, it is present in other organisms, which share a common ancestor with eukaryotes in an even more remote past, in particular in the bacterial genera *Thermus*,

*Deinococcus, Escherichia, Agrobacterium, Mesorhizobium, Helicobacter, Ureaplasma, Bacillus, Thermotoga, Synechocystis, Aquifex, Chlamydia, Mycoplasma, Mycobacterium* and *Treponema*, and the archaeal genera *Sulfolobus*, *Aeropyrum*, *Methanosarcina* and *Halobacterium*.

The orthodox way in molecular biology to interpret the almost completely universal appearance of some characteristic in a class of organisms, is that their common ancestor already did have this characteristic. The sequence NADFDGD is found in the mRNA-synthesizing RNA polymerase of all living cellular organisms, and is thus assumed to be present in their common ancestor, which lived approximately 3.5 billion years ago. This is thus the age of the sequence NADFDGD, or at least its minimum age given that there is still a long history from the origin of life up to the time of LUCA.

The second motif, with sequence GGDYDGD, corresponds to the active site of a cell-encoded RNA-dependent RNA polymerase of *Neurospora crassa,* which is part of the RdRP family according to the Pfam database. Its role is to produce double-stranded RNA from aberrant single-stranded RNA as part of an RNA silencing response (Salgado *et al.* 2006). Although this kind of protein is not present in insects, vertebrates and the yeast *Saccharomyces cerevisiae*, it is a kind of protein that is a normal constituent of an eukaryotic cell. For example, related sequences are known from the nematode worm *Caenorhabditis elegans*, the plants *Arabidopsis thaliana*, *Nicotiana tabacum* and *Oryza sativa*, and the unicellular eukaryotes *Dictyostelium discoideum* and *Giardia intestinalis* (Iyer *et al.* 2003). The GGDYDGD motif is not totally conserved among these sequences, but has the following variations: G(G/S/A)D(Y/F/M/L)DGD, where the '/' symbol means 'or'; it is thus very close to the NADFDGD motif.

The catalytic domains of these two types of RNA polymerases, Rpb1 domain 2 and RdRP, are structurally similar, whereas their other domains are completely different. The geometries of their active sites bound to $Mg^{2+}$ or $Mn^{2+}$ ions are particularly alike: they present a root mean square (rms) deviation of 0.67Å (see Table 2), and their resemblance is clearly illustrated in Fig.1. This leads us to merge these two motifs and to define a single RNA polymerase motif, of consensus sequence D(F/Y/M/L)DGD, with an invariant Gly and three invariant Asp residues. Note that the only non abundant prebiotic residue of the motif, F/Y/M/L, is not well conserved. Moreover, it points towards the protein core rather than to the ions. It has thus probably a structural role, but is unimportant for the catalytic function; this leads to the conjecture that this position was occupied by an abundant prebiotic residue A, G or V in early times. The three conserved Asp's of the motif are involved in the binding of one or two $Mg^{2+}$ or $Mn^{2+}$ ions, which coordinate the phosphates of the nucleotide triphosphate (NTP) that is going to be built into the RNA polymer, as illustrated in Figs. 2a-b.

What is the evolutionary relationship between these two RNA polymerase active sites, which share the same geometry, the same mechanism and the same Asp residues and $Mg^{2+}$ or $Mn^{2+}$ ions? Iyer *et al.* (2003) tend to view these common characteristics as a sign of very ancient common ancestry, dating from more than 3.5 billion years ago. In this view, the RNA-dependent RNA polymerase would originally have been the mRNA-producing enzyme of an organism that lived before the last common ancestor of all contemporary cells, and would have found its way to the eukaryote by a pathway of horizontal gene transfer.

The Asp residues and the Gly residue in the ancestral RNA polymerase consensus sequence xxDxDGD can be expected to play such an important role in the polymerase function that they cannot be changed, Asp because of its specific interaction with the metal ions, and Gly because of its ability to make main chain turns impossible with other amino acids. This does not exclude other ways to make an RNA polymerase; however, when starting from this particular sequence, function is lost upon mutation. Note that the extreme conservation of the other aa's in the NADFDGD motif of DNA-dependent RNA polymerase cannot be attributed to the conservation of the function, but is probably related to their interaction with other parts of the protein and other parts of the transcription machinery. Following this line of thought, xxDxDGD can be viewed as one of the most profound molecular fossils in life, and possibly one of the first coded peptide sequences. In these very early times, more than 3.5 billion years ago, they were possibly built from Val, Ala, Asp and Gly residues only, and might have been originally composed of amino acids with a prebiotic rather than a biogenic origin.

The last two DxDxD motifs presented in Table 2 correspond to the two sequences DGDAD and DGDGD, bound to $Mg^{2+}$, $Zn^{2+}$, or $Ni^{2+}$ ions. They have very similar structures, as shown in Fig. 1b; their rms deviation is equal to 0.77Å. They are also part of the same homologous family according to Pfam (Finn *et al.* 2008), named PGM_PMM_II, with consensus sequence DGD(G/A/F)D. These motifs correspond to the active sites in enzymes with mutase activity (see Fig 2.c). These enzymes can usually bind several metal ions, but $Mg^{2+}$ yields systematically the maximum activity (Regni *et al.* 2002). Some, like rabbit phosphoglucomutase, show a slight activity with $Ni^{2+}$ (Regni *et al.* 2002), whereas $Zn^{2+}$ is usually observed to inhibit the enzyme (Regni *et al.* 2006a). As an example, phosphomannomutase/ phosphoglucomutase of the bacterium *Pseudomonas aeruginosa* is changing the position on which a phosphorylated hexose sugar molecule is carrying the phosphate group. The $Mg^{2+}$ ion, which is essential for catalysis (Shackleford *et al.* 2006), is coordinated by the three Asp's. The catalytic process, in which a biphosphorylated sugar intermediate has to engage in a dramatic, 180º reorientation, is described in detail in Regni *et al.* (2006b). Of course, a complex reaction mechanism like this is a long way from simple prebiotic circumstances. Nevertheless, the proteins of the α-D-phosphohexomutase enzyme superfamily could have inherited the simple DGD(G/A)D $Mg^{2+}$ ion-binding motif from a more remote past, possibly from a very early biotic age, where this motif fulfilled simplified catalytic activities.

The DxxxxDxD motif is represented in our search by just one sequence, DAKVGDGD, which binds to two $Mg^{2+}$ ions and turned out to be the active site of dihydroxyacetone (DHA) kinase (Fig. 1c). Because this enzyme was producing the glycolytic intermediate dihydroxyacetone-phosphate (DHAP), we consider this enzymatic activity as interesting from the viewpoint of the transition of prebiotic to early biotic. Like the situation we came across with the RNA polymerases, the active site consists of two $Mg^{2+}$ ions, coordinated by three Asp residues, and these Asp residues and $Mg^{2+}$ ions are essential for catalysis (Siebold *et al.* 2003). As with the RNA polymerase active sites, the phosphate groups of an NTP are complexed with the $Mg^{2+}$ ions (Fig. 2d). The γ-phosphate is used to phosphorylate the substrate of this kinase enzyme. The substrate spectrum may be broader than the name DHA kinase suggests, because the activity of flavin mononucleotide (FMN) cyclase was found to be identical to that of human DHA kinase (Erni *et al.* 2006).

The form of the active site of DHA kinase is different from that of the RNA polymerase and mutase classes (Fig. 1). The catalytic action presents similarities with that of RNA polymerases, but differs when looking in detail: in this case the bond between the second and the third phosphate is broken, while in the case of the RNA polymerases, it is the bond between the first and the second phosphate that is being broken.

Though the three binding motifs D(F/Y/M/L)DGD, DGD(G/A)D and DAKVGDGD have different sequences and structures, they share a common submotif, DGD, with the two Asp's binding the metal ion. The conformation adopted by this submotif differs according to their host motifs, which is not surprising as such a small peptide is necessarily flexible. However, DGD could be assumed to be the ancestor of all three motifs, which could have evolved over time to fulfill more specific biological functions.

To assess the statistical significance of finding these sequence motifs in present-day proteins and proposing them as prebiotic peptide candidates, we compute the number $E$ of such motifs that are expected to be found in the PDB by chance, as defined by eqs (1)-(2). We found $E=30$, whereas the number of motifs indeed present in the PDB is equal to $N=67$ (Table 2); the associated *P-value* (see eq. (5)) is $4 \cdot 10^{-9}$. When taking into account out all biases, in particular the presence of homologous proteins in the PDB, the geometrical constraints that tend to prevent successive residues to be bound to an ion, and the preference for an Asp to be bound to an ion rather than to be close but not bound to it, we find using eqs (3)-(4) that the expected number of occurrences is $E^{cor}=1.8$, whereas its actual number is $N^{cor}=5$. Thus, we found 2.8 times more prebiotic peptide candidates than expected by chance, leading to a *P-value* of $3 \cdot 10^{-2}$. We would moreover like to stress that all these peptides are part of the active site of their host protein, although this was not among the selection criteria. This strongly improves the statistical significance of our results, even though this improvement cannot be estimated quantitatively at the moment.

As described in the state-of-the-art section, it is not easy to determine exactly which were the most abundant aa's on the primitive earth, and therefore our choice of Ala, Val, Asp and Gly as prebiotic aa's may give rise to criticisms. Indeed, Asp and Val do not come out in all prebiotic synthesis experiments (Table 1). Asp cannot be removed because its negative charge is indispensable for binding tightly to the metal ions, but Val could be dropped. However, this choice only removes a single motif from the prebiotic peptide candidates, *i.e.* DxxxxDxD. The Ala-Gly-Asp choice of abundant prebiotic aa's thus does not change the statistical significance level of our results by much, as we have $E^{cor}=1.3$ and $N^{cor}=4$, with a *P-value* of $5 \cdot 10^{-2}$.

As an additional test of our procedure, we investigate the effect of replacing Asp by His among the abundant prebiotic aa's, given that histidines also bind very efficiently to metal ions. This led to detect $N^{cor}=2$ prebiotic peptide candidates, whereas the expected number is $E^{cor}=0.1$. In these two hits, however, the structure of the host protein is incomplete: the coordinates of 5 and 11 residues, respectively, are lacking in the immediate vicinity of the metal ion. These matches have thus to be considered as spurious. Finally, when considering as prebiotic aa's all 10 aa's produced in the experiments of Ring *et al.* (1972), *i.e.* Val, Gly, Asp, Ala, Ile, Pro, Ser, Thr, Leu and Glu (Table 1), the number of expected and observed peptide candidates is of course higher: $N^{cor}=22$ and $E^{cor}=20$. However, given that this set of aa's contain the four abundant prebiotic aa's Val, Gly, Asp, and Ala, for which we found $N^{cor}=5$ and $E^{cor}=1.8$, we may conclude that the six additional aa's yield less peptide candidates than expected by chance.

In conclusion, the most statically significant results come out when considering Gly, Ala and Asp, and possibly Val, as abundant prebiotic aa's. This can be viewed as an *a posteriori* indication that these are really the true abundant prebiotic aa's.

## Discussion: the origin of life and the first peptides

How do our results fit into a wider picture of the appearance of life? The present-day picture of how life appeared is punctuated by a number of established facts, in a background of unknown or dubious stages and links. We make here an attempt to review the results that are well verified or supported, in their tentative order of occurrence:

(1) Many essential molecules for life can be synthesized in a prebiotic environment: these include amino acids (Miller 1987), but also pentoses (Ricardo *et al.* 2004), nitrogenous bases (Cleaves *et al.* 2006), and vesicle-producing organic compounds (Deamer 1985).

(2) Still in a prebiotic environment, more complex molecules can be built from these basic building blocks: adenine and ribose can be condensed into a nucleoside by the presence of borate in aqueous solution (Prieur 2001); polyphosphate, likely to be a source of energy and phosphate for producing nucleotides, can be produced by volcanic processes (Yamagata *et al.* 1991); montmorillonite clay has been discovered to catalyze oligonucleotide formation (Ferris 2006). Of particular importance to the present article, dipeptides can be produced in a dehydrating salt solution (Schwendinger & Rode, 1989), and longer peptides can be synthesized in the presence of montmorrillonite (Rode *et al.*, 1999).

(3) Vesicles, the ancestors of cells, have been shown to have a prebiotic origin (Deamer 1985) and to be catalyzed by montmorillonite (Ferris 2006). The fact that vesicles containing oligonucleotides have a higher fitness as compared with empty vesicles (Chen *et al.*, 2004), and the fact that vesicles have a ribose-collecting power (Sacerdote & Szostak, 2005) are indications of the early involvement of vesicles in the process of the origin of life. Also important is the fact that a membrane potential is not necessarily coupled to cells, but can be found in much simpler vesicles (Chen & Szostak 2004).

(4) Homochiral molecules were selected out of the racemic mixture, by a mechanism which is still not elucidated.

(5) At some stage RNA and peptides became part of the same biochemical system. However, which of the RNA or peptides came first is highly controversial. The arguments in favor of the RNA world include the existence of ribozymes, the essential role of RNA in transcription and translation, and the fact that aa's are activated by ATP before they are coupled to their cognate tRNAs. An argument against the RNA world is the huge size of the RNA-dependent RNA-polymerizing ribozyme (Johnston *et al.*, 2001). The arguments in favor of the peptide world include the fact that peptides are more easily synthesized and more resistant to degradation (Plankensteiner *et al.*, 2005). Moreover some peptides have been shown to catalyze their own synthesis (Lee et al., 1996) and have catalytic activity (Kochavi *et al.*, 1997). A hybrid picture between these two extremes would be an RNA-peptide world, in which both types of molecules evolved together.

(6) The coding of genetic information in DNA was probably, given the difficulty of synthesizing DNA, one of the last steps before the appearance of life as we know it. But this issue is far from settled.

Exactly how these different more or less well established facts are related is the subject of much debate and speculation, and is outside the scope of the present paper. Here we addressed the issue of the form and functionality of the first peptides. We believe it should be possible to answer the latter question to a large extent independently of how points (1) to (6) fit together. It can also to some extent be addressed independently of how the first peptides were synthesized and how their sequences were coded, and of whether a peptide, RNA, or RNA-peptide world appeared first. Concerning the last point there are two conflicting results which cannot be resolved at present: on the one hand small peptides can be synthesized in prebiotic environments as mentioned in point (2) above, but on the other hand peptide synthesis as it occurs in the cell is deeply connected to the RNA world, and the traces of this connection can still be seen in the essential role in translation of mRNA, rRNA, and tRNA. However this may not have been the primitive situation.

Two main facts can be stated with confidence about the first functional peptides: first of all the first peptides were presumably made of the amino acids that are efficiently produced in prebiotic synthesis, as these are the chemically simplest aa's. A supporting argument is that the present genetic code, with its block structure, could easily have evolved from a more primitive code that used fewer amino acids: from time to time some codons were reassigned to other aa's without affecting the other codons.

Secondly, the first peptides probably appeared when a coding mechanism for the aa sequence was inexistent, or at least highly imperfect and error prone. Indeed the present (highly complex) coding, with its essentially error free mechanisms for duplication, transcription and translation, was almost certainly not present initially, but presumably evolved in response to the increasingly critical role of complex proteins in early life. This is supported by the fact that the genetic code seems to have been optimized against transcription and translation errors (Freeland & Hurst, 1998)

A first question to address when trying to understand what were the first functional peptides is to decide whether they had lengths comparable to present-day protein chains (about 20 to $10^4$ aa's long), or whether they were much shorter. We believe that the first possibility is highly unlikely. Indeed long random peptides will not have a well defined 3-dimensional structure and certainly will not have any useful functionality. On the other hand random synthesis of very short peptides could have produced useful ones with reasonably high yield. For instance the common submotif of our prebiotic peptide candidates, DGD, could be one of those randomly selected peptides. In this case the success rate of the synthesis would have been $4^{-3}=1/64$ (assuming that only the four abundant prebiotic residues were used and were equally abundant). The fraction of active versus possible peptides drops very rapidly as the length increases (at least if all positions have to be occupied by specific residues): for 8 residues it is as small $4^{-8}$. However, if the synthesis occurred in the presence of metal ions, which tend to attract free Asp's, the amount of meaningful peptides could be much higher. The prebiotic active sites of 5-8 residues long, such as those we identified in this paper, seem to form the limit between randomly synthesized and template-directed peptides.

To discover what the original functional peptides were, we have adopted here a theoretical approach. We have identified, using the method described above, active sites of present-day proteins that are made of prebiotic amino acids, carry out functionalities that are essential for early life, and can be traced back to LUCA.

We then take the bold step of supposing that these sequences in fact go back much earlier, and constitute some of the earliest useful peptides. This is of course highly conjectural. If we accept this conjecture, we obtain a remarkably precise picture of the chemical role, the first, very short, oligopeptides could have played during very early life. In all cases we have identified, phosphate groups are manipulated, mainly by $Mg^{2+}$ ions, coordinated by oligopeptide motifs consisting of three Asp residues and at least one Gly residue. Remarkable is the fact that ribozymes also need specifically $Mg^{2+}$ ions for both structure and function (Glasner *et al.*, 2002; Takagi *et al.*, 2004; Robertson & Scott, 2007). $Mg^{2+}$ ions seem thus to have a particularly important role in handling phosphate groups and therefore in the development of life.

Moreover, in the peptides we identified, the biological functions performed by these active sites, *i.e.* RNA polymerization, glycolytic intermediate production and sugar phosphorylation, can be assumed as prebiotic, or at the very least, very early biotic. The existence of these early peptides would of course have required the existence of a stable environment, protected from toxic chemical species. This environment could have been provided by the vesicles that correspond to the ancestors of cells (point (3) above). At present we do not have enough information to decide whether these oligopeptide motifs appeared before, during or after the RNA world. In the latter case they could have constituted an interesting improvement to the phosphate chemistry available in the purely RNA world.

More generally, we can conjecture that, once small random peptides became useful to early life, mechanisms emerged to select for the active peptides. Thus gradually a coding and translation mechanism appeared, which in turn would allow for the synthesis of longer peptides, and for the exploration of the space of possible aa sequences, progressively selecting more active and specific oligopeptides and proteins. Therefore, the appearance of prebiotic oligopeptides is tightly connected with the appearance of the genetic code. Note however that we have no idea about how the transition occurred from the production of interesting compounds by noncoded synthesis of random peptides to the production of coded peptides by a ribosome, and about the gradual evolutionary construction of the complex system of tRNAs, large and small rRNAs and mRNAs; all this occurred moreover in the presence of the various toxic substances that must have existed on a primeval earth. These issues, however, are beyond the scope of the present paper.

## Conclusion: how to validate our findings?

We identified in nowaday protein structures five short segments bound mainly to $Mg^{2+}$ ions, and built for 80% at least from the four aa's Asp, Gly, Val, Ala, thought to be the most abundant aa's in prebiotic times. These motifs appear as statistically significant as they are 2.8 times more frequent than expected by chance, with a *P-value* of $3~10^{-2}$. Moreover, they all form the active sites of their host protein: DFDGD and DYDGD in RNA polymerases, DGDGD and DGDAD in mutases, and DAKVGDGD in dihydroxyacetone kinase, where they manipulate phosphate groups, thought to be an important biological function in the very first stages of life. We thus conjecture that these motifs could correspond to the first functional peptides, at the earliest stages of life, before or at the beginning of the setup of the genetic coding mechanism, and after the purely RNA world or during the RNA-peptide world. These peptides could be viewed as transitional fossils, constituting a missing link between the prebiotic amino acids and the coded proteins.

Note that this conjecture must considered with caution, as there could be alternative explanations for the prominent presence of Ala, Val, Asp and Gly in small, metal-binding, universal active sites. For example, these aa's could be indispensable for achieving a specific biological function. This argument is however dubious, given that other RNA polymerases are known to exist with active sites consisting of different aa's. Another explanation is that an intense process of selection within the first organisms led to the small, metal-binding active sites notwithstanding the presence of many more proteinaceous aa's in the prebiotic environment and other possible ways of making the active site.

How could our suggestions be tested experimentally? The easiest point to analyze is whether the short prebiotic peptide candidates are soluble in aqueous solution, and whether they form more or less unique and stable structures with $Mg^{2+}$ ions. The second point to test is whether these short, metal ion-complexed peptides do still have some kind of enzymatic activity, even very inefficient, when they are isolated from their normal protein context. This would possibly require introducing other environments, present in the prebiotic era, which would favor the assembly of the different partners required for enzymatic function. Finally, we could consider peptides built from the abundant prebiotic residues only, for example ADVDGD and DAAVGDGD, which could represent even earlier peptides, and test if their properties are comparable to those of ADFDGD and DAKVGDGD. A further area of research is the interaction of these kinds of oligopeptides with RNA molecules.

In summary, the present work pursues a new approach to studying very early life. Through the analysis of the sequence and structure of present-day proteins, more and more of which are available in public databases, we try to push the well established method of phylogeny to even remoter periods. In this way we formulate new hypotheses about the origin of life.

## Acknowledgments

We thank Roman Laskowski for his kind help with PDBsum. MR and DG acknowledge support from the Belgian State Science Policy Office through an Interuniversity Attraction Poles Program (DYSCO); MR, DG and SM acknowledge support from the Fund for Scientific Research (FRS) through FRFC projects; PvdG and HB acknowledge support from the Bsik research grant BRICKS, and HB from an NWO VICI grant. SM and MR are Research Directors at the FRS.

|  |  | Gly | Ala | Asp | Val | Ser | Glu | Other |
|---|---|---|---|---|---|---|---|---|
| **Prebiotic synthesis** | Miller and coworkers[1] | 1 | $9 \cdot 10^{-3}$ - 2 | $2 \cdot 10^{-4}$ - $8 \cdot 10^{-2}$ | 0 - $4 \cdot 10^{-2}$ | $1 \cdot 10^{-4}$ - $3 \cdot 10^{-2}$ | 0 - $2 \cdot 10^{-2}$ | 0 - $4 \cdot 10^{-2}$ |
|  | Plankensteiner *et al.* (2006) | yes | yes | 0 | yes | ? - yes | 0 | yes |
| **Meteorites** | Botta *et al.* (2002) | 1 | 0 - $7 \cdot 10^{-1}$ | $6 \cdot 10^{-2}$ - $6 \cdot 10^{-1}$ | ? - yes | 0 - $5 \cdot 10^{-1}$ | $2 \cdot 10^{-2}$ - 2 | ? |

Table 1. Relative frequencies of proteinaceous aa's in prebiotic synthesis experiments and in meteorites. [1]Miller (1955); Ring *et al.* (1972); Schlesinger & Miller (1983); Johnson *et al.* (2008).

| Binding motif[1] | Candidate prebiotic sequence[1] | Metal ion | $N$ | $N^{cor}$ | Biological function | Central structure[2] | Mean Rms[3] (Å) | Rms[4] (Å) | Consensus sequence[5] |
|---|---|---|---|---|---|---|---|---|---|
| **DxDxD** | **AD**F**DGD** | Mg, Mn | 53 | 1 | DNA-directed RNA polymerase Rpb1 domain 2 | 1i6h (A480-A485) | 0.63 | 0.67 | **D**(F/Y/M/L)**DGD** |
|  | GG**D**Y**DGD** | Mg | 4 | 1 | RNA-dependent RNA polymerase | 2j7n (A1005-A1011) | 0.03 |  |  |
|  | **DGDGD** | Mg, Zn, Ni | 7 | 1 | Phosphoglucomutase, phosphomannomutase | 1p5d (X242-X246) | 0.45 | 0.77 | **DGD**(G/A/F)**D** |
|  | **DGDAD** | Zn | 1 | 1 | Phosphoglucomutase | 1kfi (B308-B312) | - |  |  |
| **DxxxxDxD** | **D**AKV**GDGD** | Mg | 2 | 1 | Dihydroxyacetone kinase | 1un9 (A380-A387) | - | - | **DAKVGDGD** |

Table 2. Prebiotic peptide candidates found in the protein structure database. [1]The residues in bold are bound to the metal ion. [2]The central structure of the group is that for which the rms deviation with respect of all other members of the group is minimum. [3]The rms deviation is computed between all heavy atoms for the main chain and side chain and the metal ion, after coordinate superimposition; the mean is taken between all pairs in the class. [4]The rms deviation is computed between the equivalent parts of the central structures of two classes, in particular between DFDGD and DYDGD (considering the common atoms in F and Y), and between DGDGD and DGDAD (considering the common atoms in G and A). [5]The consensus sequence is obtained from sequence alignments in homologous protein families.

# Figure legends

**Figure 1.** Three types of metal ion binding prebiotic peptide candidates. The figures are drawn using PyMol (http://pymol.sourceforge.net). (a) D(F/Y)DGD motif in RNA polymerases Yellow: PDB entry 1i6h A480-A485 bound to a $Mg^{2+}$ ion; blue: PDB entry 2j7n A1005-A1011 bound to a $Mg^{2+}$ ion. (b) DGD(G/A)D motif in phospho(gluco/manno)mutase. Yellow: PDB entry 1p5d X242-X246 bound to a $Zn^{2+}$ ion; blue: PDB entry 1kfi B308-B312 bound to a $Zn^{2+}$ ion. (c) DAKVGDGD motif in dihydroxyacetone kinase. Blue: PDB entry 1un9 A380-A387 bound to two $Mg^{2+}$ ions.

**Figure 2.** Biological function of the three prebiotic peptide candidates. The figures are drawn using PyMol (http://pymol.sourceforge.net). (a) DFDGD motif: PDB entry 1i6h A480-A485 bound to a $Mg^{2+}$ ion (blue), interacting with an RNA strand (red), itself in interaction with a DNA strand (pink). (b) DFDGD motif: PDB entry 2e2h A480-A485 bound to two $Mg^{2+}$ ions (blue), interacting with a GTP molecule (orange) and an RNA strand (red), itself in interaction with a DNA strand (pink). (c) DGDGD motif: PDB entry 1p5d X242-X246 bound to a $Zn^{2+}$ ion (yellow), and close to a G1P molecule (red). (d) DAKVGDGD motif: PDB entry 1un9 A380-A387 bound to two $Mg^{2+}$ ions (blue) and interacting with an ATP molecule (red).

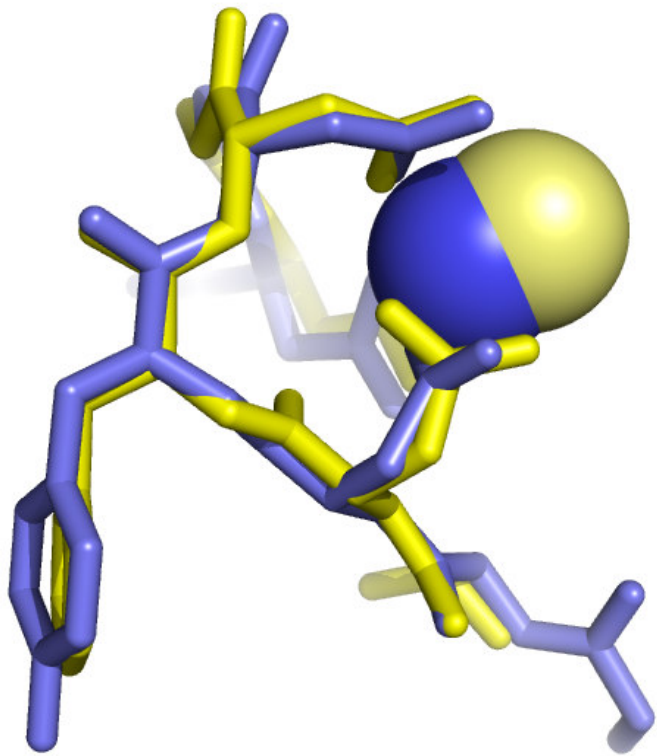

Figure 1.a

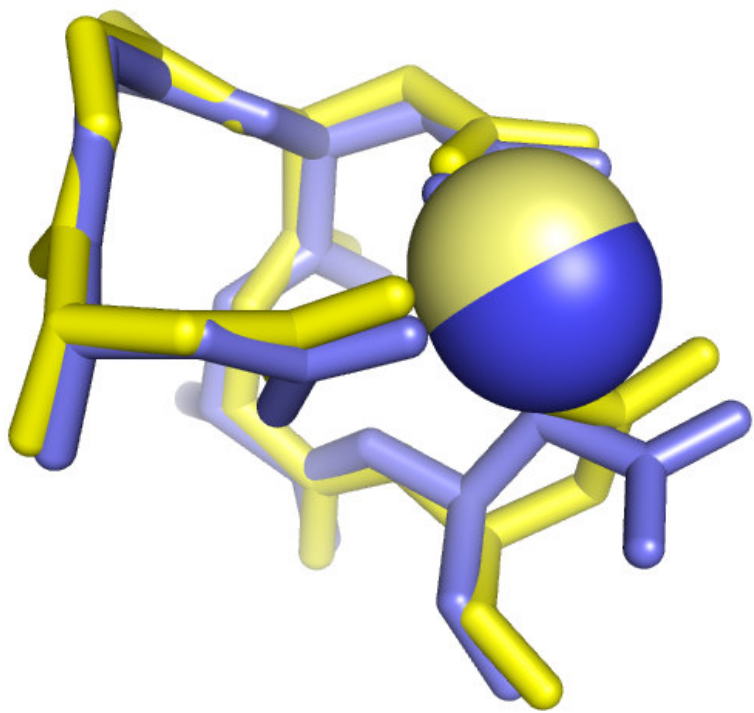

Figure 1.b

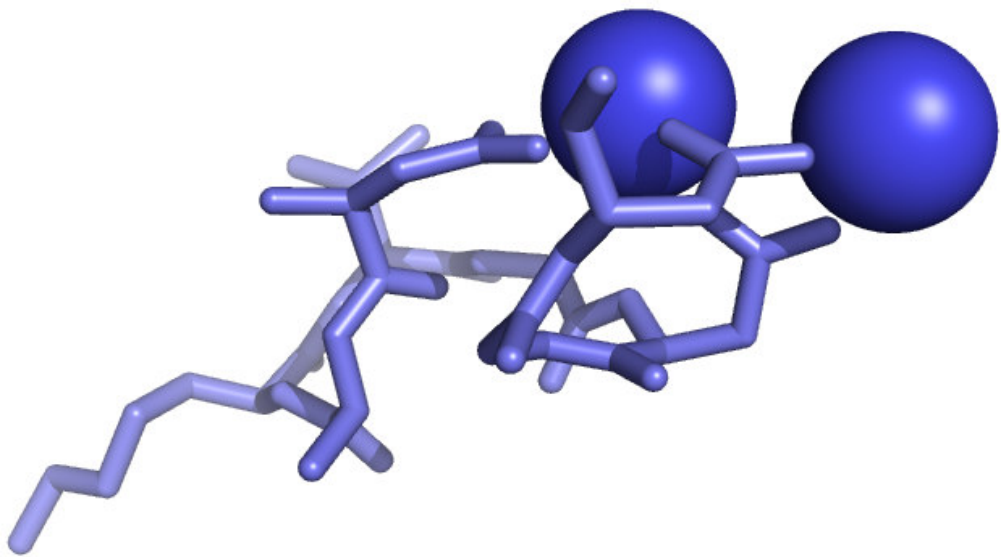

Figure 1.c

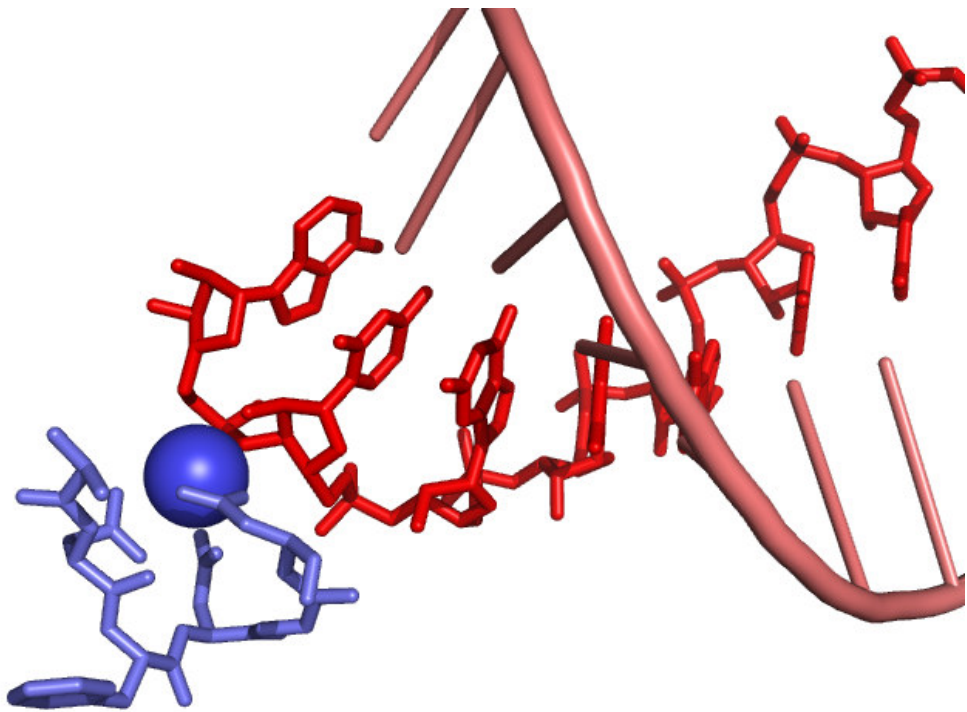

Figure 2.a

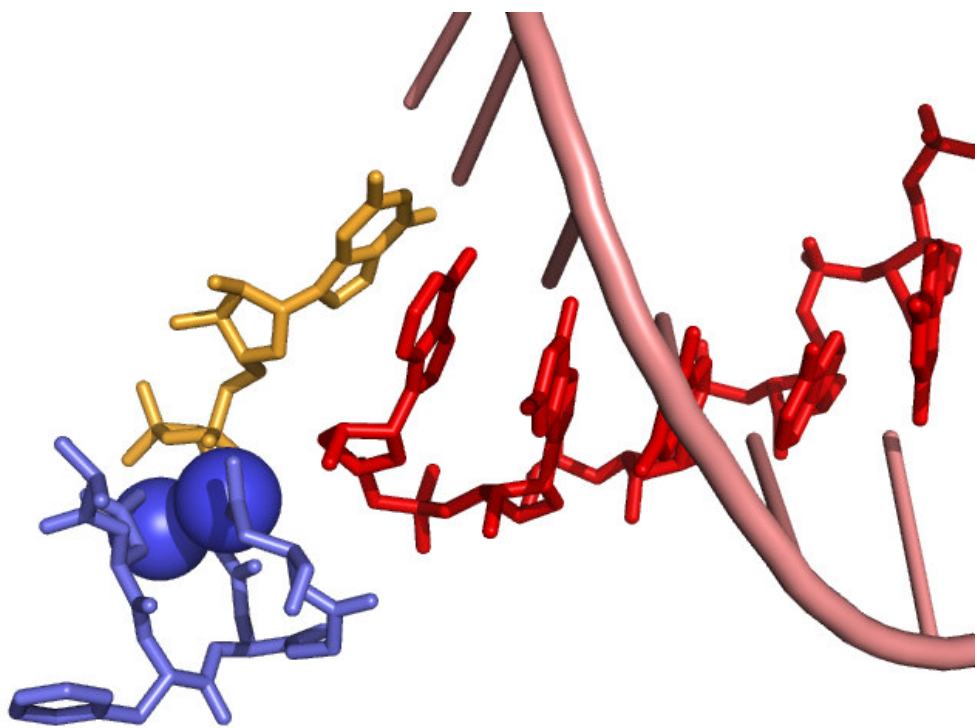

Figure 2.b

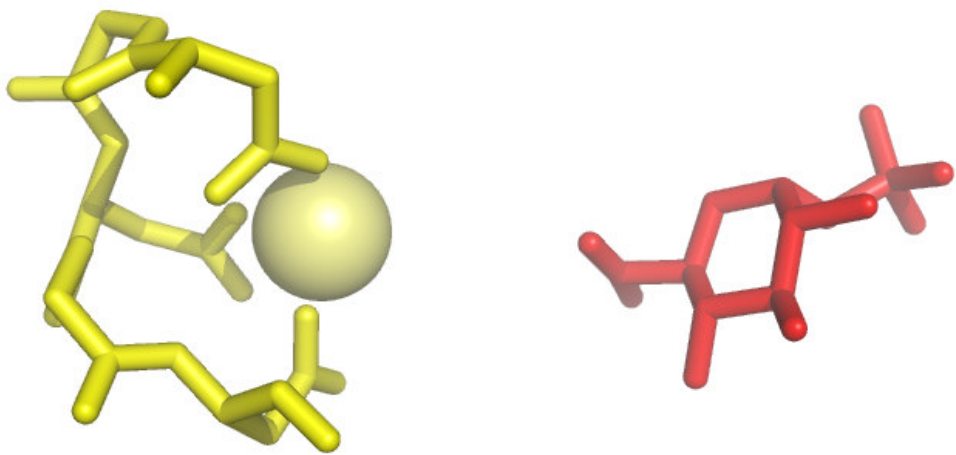

Figure 2.c

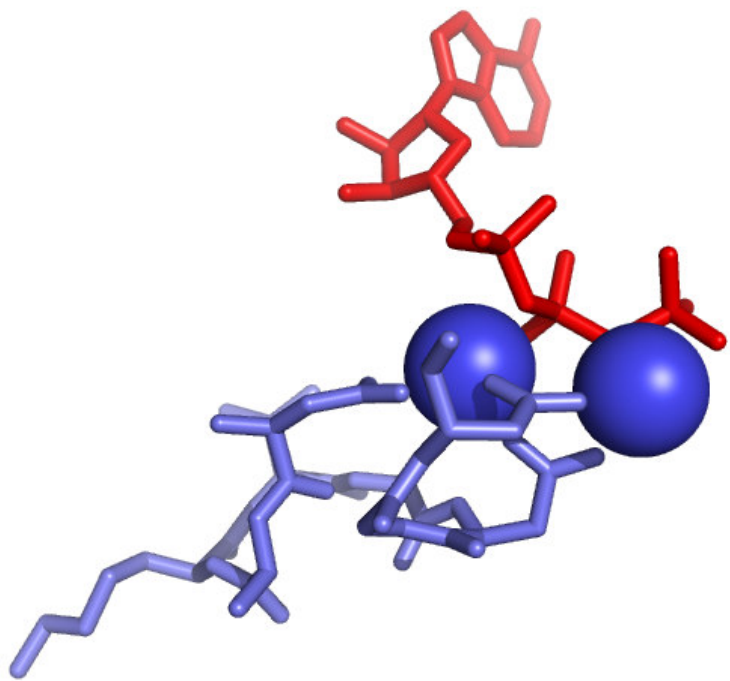

Figure 2.d